\documentclass[twocolumn,showpacs,preprintnumbers,amsmath,amssymb,prl,superscriptaddress,floatfix]{revtex4}

\usepackage{graphicx}
\usepackage{dcolumn}
\usepackage{bm}
\usepackage{color}
\usepackage{xspace}

\newcommand{\Msun}{\ensuremath{{\rm M}_{\scriptscriptstyle \odot}}\xspace}

\begin{document}

\preprint{APS/000-000}

\title{Neutrino Nucleosynthesis}


\author{A.~Heger}
 \affiliation{Department of Astronomy and Astrophysics, The University of Chicago, 5640 S. Ellis Ave, Chicago, IL 60637, USA}
 \affiliation{Theoretical Astrophysics Group, MS B227, Los Alamos National Laboratory, Los Alamos, NM 87545, USA}
\author{E.~Kolbe}
 \affiliation{Departement f\"ur Physik, Universit\"at Basel, Basel, Switzerland}
\author{W.~C.~Haxton}
 \affiliation{Institute for Nuclear Theory, University of Washington, Seattle 98195, USA}
\author{K.~Langanke}
 \affiliation{Institut for Fysik og Astronomi, {\AA}rhus Universitet, DK-8000 {\AA}rhus C, Denmark}
\author{G.~Mart\'{\i}nez-Pinedo}
\affiliation{Institut d'Estudis Espacials de Catalunya, Edifici Nexus, Gran Capit\`a 2, E-08034 Barcelona, Spain}
\affiliation{Instituci\'o Catalana de Recerca i Estudis Avan\c{c}ats, Llu\'{\i}s Companys 23, E-08010 Barcelona, Spain}
\author{S.~E.~Woosley}
 \affiliation{Department of Astronomy and Astrophysics, University of California, Santa Cruz, CA95064, USA}
\date{\today}

\begin{abstract}
We study neutrino process nucleosynthesis in massive stars using newly
calculated cross sections, an expanded reaction network, and complete
and self-consistent models of the progenitor star.  We reevaluate the
production of light isotopes from abundant progenitors as well as that
of rare, heavy, proton-rich isotopes.  In particular, new results are
given for $^{11}$B, $^{19}$F, $^{138}$La, and $^{180}$Ta.  The
production of these isotopes places limits on neutrino spectrum and
oscialltions.
\end{abstract}


\maketitle

Nuclei can be synthesized in the mantle of a core collapse supernova
by the neutrino process \cite{Dom78} -- energetic supernova neutrinos
excite nuclei above particle breakup through neutral- and
charge-current reactions, creating new daughter nuclei. While
typically only $1\,\%$ of mantle nuclei experience inelastic neutrino
reactions, certain rare nuclei, one mass unit below abundant parent
nuclei such as C, O, and Ne, may nevertheless be produced dominantly
by the neutrino process.  The final abundance depends not only on the
instantaneous yield of the daughter isotope, but also whether that
isotope survives subsequent processing.  
In many cases the heating associated with the passage of the shock
destroys the daughter isotope: the surviving nuclei may be only those
produced post-shock, after the relevant shell has expanded and cooled.

Two recent developments make a re-examination of the $\nu$-process
timely.  First, we now know about neutrino oscillations, which could
alter certain $\nu$-process yields by enhancing charged-current
production channels.  Second, new data on the abundances of B and F --
two key $\nu$-process products -- have been obtained.  Prochaska,
Howk, and Wolfe \cite{pro03} recently observed over 25 elements in a
galaxy at redshift $z=2.626$, whose young age and high metallicity
implies a nucleosynthetic pattern dominated by short-lived, massive
stars.  Their observation of a solar B/O ratio in an approximately
1/3-solar-metallicity gas argues for a primary (metal-independent)
production mechanism, such as the $\nu$-process (making $^{11}$B),
rather than a secondary process, such as cosmic ray proton spallation
reactions on interstellar CNO seed nuclei (making $^{10}$B and
$^{11}$B).  The new F abundance data of Cunha et al.\ \cite{cunha03}
showing a low F/O ratio in two $\omega$ Centauri stars argue against
AGB-star production of F, and are quite consistent with $\nu$-process
models (though also with production in cores of stars sufficiently
massive to be Wolf-Rayet stars at the beginning of He burning).

Until the present effort the most complete neutrino process
calculations were those of \cite{woo90}, who evaluated productions in
a $20\,\Msun$ Pop I star evolved without mass loss, including
semiconvection, and using the Caughlan {\it et al.} \cite{cau85}
$^{12}$C($\alpha,\gamma)^{16}$O reaction rate.  The nuclear chemistry
of the coproduced protons and neutrons, which proved to reduce
productions of important isotopes like $^{11}$B, $^{15}$N and
$^{19}$F, was followed in a nuclear reaction network.  The effects of
shock-wave heating and post-shock neutrino process production, in
shells expanding off the star and cooling, were evaluated.  Later
Timmes \textsl{et al.} \cite{timmes95} extended this calculation to a
full galactic model, integrating the neutrino process over a range of
progenitor stars with evolving metalicity.

This letter extends this earlier work in several important ways.  One
is the incorporation of mass loss in the evolution of the progenitor
star.  Second, for the first time a reaction network is employed that
includes all of the heavy elements through bismuth using updated
reaction rates \cite{rau02}: the work of \cite{woo90,timmes95} ranged
only up to zinc.  This allows us to add selected neutrino reactions --
the inclusion of neutrino cross sections for the entire extended
network has not yet been attempted -- for products that may serve as
electron-neutrino "thermometers."  These cross sections are evaluated
in a model that we believe is suitable for heavy nuclei.  Third, the
nuclear evaporation process -- emission of a proton, neutron, or
$\alpha$ -- is treated in a more sophisticated statistical model that
takes into account known nuclear levels and their spins and parities.

The various partial neutrino cross sections are calculated as two-step
processes, as in \cite{woo90}. The charged- and neutral-current cross
sections are evaluated as function of excitation energy in the final
nucleus, then a statistical model is used to evaluate the subsequent
decay by particle or $\gamma$ emission.  For the $p$- and $sd$-shell
nuclei the Gamow-Teller (GT) contributions to the $(\nu_e,e^-)$ and
$(\nu,\nu')$ responses were taken from $0\hbar \omega$ shell model
diagonalizations, as appropriate.  The Cohen-Kurath ($^{14}$N)
\cite{Kurath} and Brown-Wildenthal \cite{BW88} interactions were used.
For $^{12}$C we adopted energy and GT strength of the main GT
transition, to the $T=1$ state at $15.11\,$MeV in $^{12}$C or its
analogue in $^{12}$N, from experiment.  The double-magic nucleus
$^{16}$O has no GT response in the 0$\hbar \omega$ limit.  All other
contributions to the neutrino cross sections have been determined
within the random phase approximation (RPA) considering multipoles up
to $J=4$ and both parities.  The RPA model, described in \cite{KLV},
treats proton and neutron degrees of freedom separately and employs a
partial occupancy formalism 
for non-closed-shell nuclei. The residual interaction is a zero-range
Migdal force.  As realistic shell-model calculations for the heavy
nuclei ($^{138}$Ba, $^{139}$La, $^{181}$Ta, $^{180}$Hf) are yet not
practical, the entire response was calculated using RPA.

In the second step we use the statistical model SMOKER \cite{Cowan91}
to calculate, for each final state with well-defined energy, angular
momentum, and parity, branching ratios for p, n, $\alpha$
and $\gamma$ emission.  SMOKER uses experimentally determined levels in
the daughter nucleus, supplemented at higher energies by an
appropriate level density formula \cite{Cowan91}.  If the decay leads
to an excited level of the daughter nucleus above particle threshold,
the subsequent decay of this level is treated similarly.  The yield of
a given nucleus is obtained by folding the various branching ratios,
as a function of energy, with the neutrino response function.

This is qualitatively the same procedure used in \cite{woo90}.  There
full multi-shell shell-model calculations were done through $^{16}$O;
in the $sd$ shell, however, the first-forbidden response was taken
from the simpler Goldhaber-Teller model.  The $sd$ positive-parity
shell model calculations for $^{24}$Mg, $^{28}$Si, and $^{34}$S were
also truncated.  
Perhaps more important, the branching ratios were evaluated in
\cite{woo90} with a statistical model that lacked the capabilities of
SMOKER (e.g., experimental level densities and spin/parity selection
rules).  In general, the present cross sections turn out to be
slightly smaller than those of \cite{woo90}.  Detailed partial cross
sections for the heavier nuclei ($^{138}$Ba, $^{139}$Ta, $^{180}$Hf,
$^{181}$Ta) have not been previously calculated.


\begin{figure}[htb]
\caption{Dominant neutrino process cross sections for production for
$^{11}$B, $^{19}$F, $^{139}$La, and $^{180}$Ta as a function of
neutrino temperature (degeneracy parameter $\alpha=0$).}
\label{fig:cs}
\includegraphics[width=\columnwidth,clip]{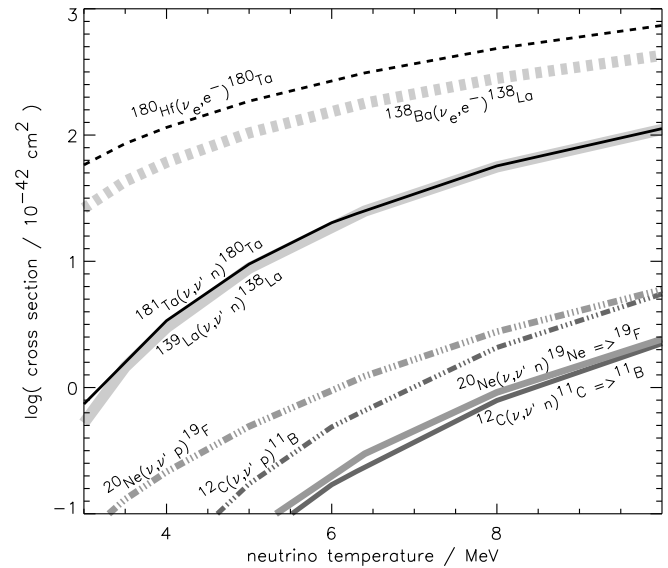}
\end{figure}

\begin{figure}[htb]
\caption{Electron neutrino charged current cross sections on
$^{138}$Ba for cascades up to two particle emission.}
\label{fig:csba138}  
\includegraphics[width=\columnwidth,clip]{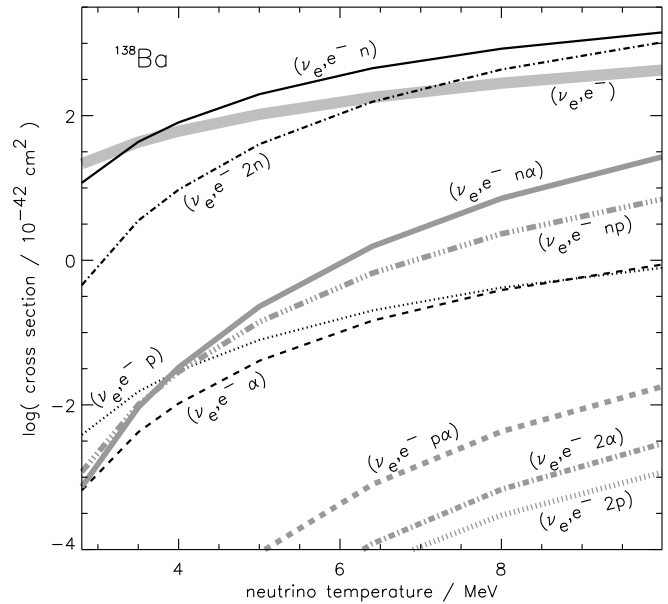}
\end{figure}


For the neutrino spectra we took Fermi-Dirac distributions with zero
chemical potential and temperatures $T=6\,$MeV for $\mu$ and $\tau$
neutrinos and their antiparticles and $T=4\,$MeV for $\nu_e$ and
$\bar{\nu}_e$.  Very recent supernova simulations \cite{KRJ03} find
somewhat harder $\nu_{\mu,\tau}$ spectra with $T=5.9\,$MeV and a
degeneracy parameter $\alpha=1.5$, predicting slightly larger average
neutrino energies and increased cross sections (for example by
$13\,\%$ for $^{20}$Ne($\nu$,$\nu\prime$)).  The $\bar{\nu}_e$
energies are also found to be more energetic than what we assumed, but
they have only little influence on the nuclei studied here.

Potentially important $\nu$-process candidates can be identified by
looking at the abundances in the star during the explosion, after the
passage of the shock.  This includes in particular radioactive parent
nuclei, existing in nuclear statistical equilibrium or produced in the
passage of the shock wave.  
Table~\ref{table:NuCand} shows some of the candidate reactions that
were identified from abundances found in our progenitor star $3.85\,$s
after core bounce, the time when the shock reaches the base of the
helium shell, located at radius $4 \times 10^9\,$cm and mass
coordinate $6.3\,\Msun$.  The table includes an estimate of the
neutrino cross section $\sigma_\odot$, required to produce the solar
abundance of each isotope $Y_\odot$, measured with respect to
$^{16}$O, when averaged over all ejecta: $\sigma_\odot := 4\pi Y_\odot
M_{\mathrm{ejecta}} / n_\nu \int_{\mathrm{ejecta}} Y_{\mathrm{parent}}
r^{-2} \mathrm{d}m$.  In this integration we use the progenitor radial
coordinates $r$ to locate the various shells, which would
underestimate the needed cross sections for isotopes produced
dominantly after shock wave passage (when the shell is expanding
outward).  We note the two most favorable candidates (smallest
required cross sections), $^{51}$V and $^{55}$Mn, were produced at
significant levels in the $\nu$-process work of \cite{woo90}: those
calculations used an interpolation method to generate neutrino cross
section estimates for all nuclei in the network.  (However, the
$^{51}$V and $^{55}$Mn were made as $^{51}$Mn and $^{55}$Co by the
neutral current reactions $(\nu,\nu'$p) off of $^{52}$Fe and
$^{56}$Ni, respectively, not by the charge-current reactions listed in
the table.)  The remaining reactions in Table~\ref{table:NuCand} will
be explored in future, once the needed neutrino cross sections are
evaluated.

\begin{table}[bt]
\centering
\caption{Heavy $\nu$-process candidate reactions as derived form a
25\,\Msun stellar model $3.85\,s$ after core bounce.
$\sigma_{\odot,42}$ is the cross section in $10^{-42}\,$cm$^2$ that
would be required for solar production of the isotope.%
\label{table:NuCand}} %
\begin{ruledtabular}
\begin{tabular}{r@{$\!\!\!\!\!\!$}lrr@{$\!\!\!\!\!\!$}ll}
\multicolumn{2}{c}{product} &
\multicolumn{1}{c}{$\sigma_{\odot,42}$} &
\multicolumn{2}{c}{parent} &
\multicolumn{1}{l}{process} \\
\noalign{\smallskip}
\hline
\noalign{\smallskip}
$^{ 51}$&V  &     3.0 & $^{ 52}$&Fe  & ($\nu_e$,$e^-\,$n) \\
$^{ 55}$&Mn &     2.0 & $^{ 56}$&Ni  & ($\nu_e$,$e^-\,$p) \\
$^{ 78}$&Kr &  11; 29 & $^{ 80}$&Kr  & ($\nu$,$\nu'\,$2n); ($\nu_e$,e$^-\,$2n) \\
$^{ 84}$&Sr &  19; 50 & $^{ 86}$&Sr  & ($\nu$,$\nu'\,$2n); ($\nu_e$,e$^-\,$2n) \\
$^{   }$&   &     102 & $^{ 85}$&Rb  & ($\nu_e$,e$^-\,$n) \\
$^{138}$&Ce &  33; 88 & $^{140}$&Ce  & ($\nu$,$\nu'\,$2n); ($\nu_e$,e$^-\,$2n) \\
$^{180}$&Ta &      58 & $^{182}$&Hf  & ($\nu_e$,e$^-\,$2n) \\
$^{   }$&   &     118 & $^{181}$&Hf  & ($\nu_e$,e$^-\,$n) \\
$^{   }$&   &      22 & $^{182}$&W   & ($\bar{\nu_e}$,e$^+\,$2n) \\
$^{196}$&Hg &      74 & $^{198}$&Pb  & ($\nu$,$\nu'\,$2n) \\
$^{   }$&   &     191 & $^{198}$&Pb  & ($\nu_e$,e$^-\,$2n), ($\bar{\nu_e}$,e$^+\,$2n), $\ldots$ \\
\end{tabular}
\end{ruledtabular}
\end{table}

Our full $\nu$-process calculations included all dynamics, explosive
and $\gamma$-process \cite{Aud70} nucleosynthesis, and all network
reactions destroying parent or daughter isotopes involved in
$\nu$-process synthesis. Supernova calculations were done with the
KEPLER code \cite{WZW78}, starting from the $15$ and
$25\,\mathrm{M}_\odot$ models S15 and S25 of \cite{rau02}.  We assume
a total energy of $5\times10^{52}\,$erg per species with a luminosity
exponentially decaying after onset of core collapse on a time-scale of
$3\,$s and constant neutrino temperature of $4\,$MeV for the
$\nu_{\mathrm{e}}$ and $\bar{\nu}_{\mathrm{e}}$ and of $6\,$MeV for
the $\nu_{\mu}$, $\bar{\nu}_{\mu}$, $\nu_{\tau}$, and
$\bar{\nu}_{\tau}$, with zero degeneracy parameter ($\alpha=0$).

\begin{figure}[ht]
\caption{Production of $^{138}$La in a $25\,$M$_\odot$ star.  We give
the production of $^{138}$La without neutrinos, with the charged
current reaction $^{138}$Ba($\nu_{\mathrm{e}}$,e$^-$), with the
neutral current reaction $^{139}$La($\nu$,$\nu'$n) and with both
reactions.  Additionally, we show the SN shock temperature, the pre-SN
density, and the neutrino energy fluence $F_{\mathrm{E},\nu}$ assuming
a total neutrino energy of 3$\times10^{53}\,$erg.}
\label{fig:la138}
\includegraphics[width=\columnwidth,clip]{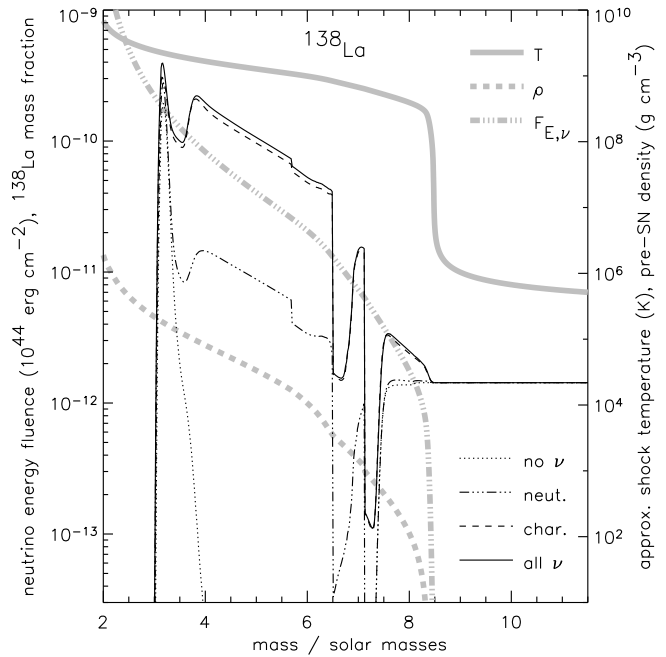}
\end{figure}

\begin{figure}[ht]
\caption{Production of $^{138}$La in a $25\,$M$_\odot$ star and its
neutrino process progenitor nuclei, $^{139}$La (neutral current) and
$^{138}$Ba (charged current).  The mass fraction of these isotopes as
a function of the enclosed mass is shown before (\textsl{gray}) and
after (\textsl{black}) the supernova explosion.}
\label{fig:la138prog}
\includegraphics[width=\columnwidth,clip]{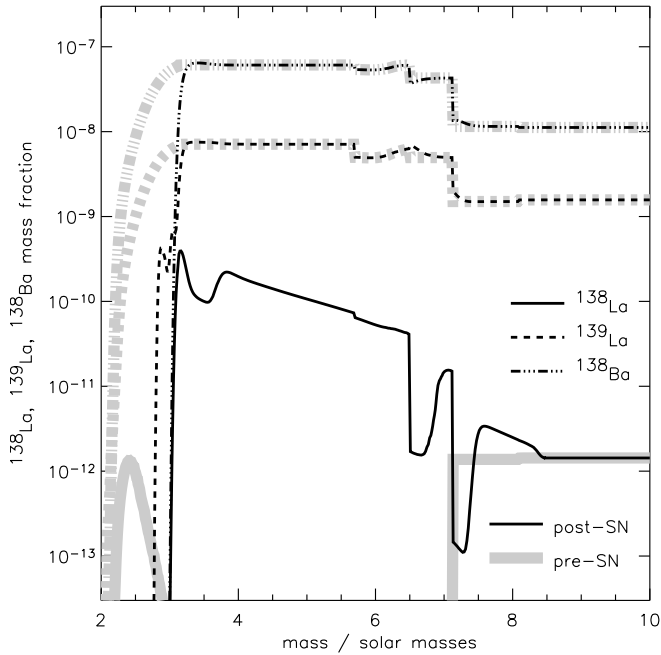}
\end{figure}



We evaluated neutral current reactions and two particle cascades for
$^{12}$C, $^{14}$N, $^{16,18}$O, $^{20,22}$Ne, $^{24,26}$Mg,
$^{28,30}$Si,$^{139}$La and $^{181}$Ta. 
For charged current reactions we include 
two particle cascades for $^{138}$Ba and $^{180}$Hf
(Figure~\ref{fig:cs}).  Figure~\ref{fig:csba138} shows that the 2n
channel is important for these nuclei.

\begin{table}
\centering
\caption{Production factor relative to solar normalized to $^{16}$O
production as a function of $T_{\!\nu_{\!\mathrm{e}}}$ (for charged
current only) and using 6$\,$MeV for the $\mu$ and $\tau$ neutrinos. %
\label{table:cc}}
\begin{ruledtabular}
\begin{tabular}{lr@{$\!\!\!\!\;$}lrrrrr}
\multicolumn{1}{c}{star} &
\multicolumn{2}{c}{product} &
\multicolumn{1}{r}{(no $\nu$)} &
\multicolumn{1}{r}{(no $\nu_{\mathrm{e}}$)} &
\multicolumn{1}{r}{$4\,$MeV} &
\multicolumn{1}{r}{$6\,$MeV} &
\multicolumn{1}{r}{$8\,$MeV} \\
\noalign{\smallskip}\hline\noalign{\smallskip}
&$^{ 11}$& B &  0.011 &  1.509 &  1.899 &  3.291 &  ----- \\
&$^{ 15}$& N &  0.396 &  0.480 &  0.486 &  0.530 &  ----- \\
15\,M$_\odot$ 
&$^{ 19}$& F &  0.375 &  0.577 &  0.643 &  0.914 &  ----- \\
&$^{138}$&La &  0.190 &  0.279 &  0.974 &  1.734 &  2.456 \\
&$^{180}$&Ta &  0.599 &  1.016 &  2.751 &  4.628 &  6.026 \\
\noalign{\smallskip}\hline\noalign{\smallskip}
&$^{ 11}$& B &  0.004 &  0.828 &  1.170 &  2.384 &  ----- \\
&$^{ 15}$& N &  0.039 &  0.112 &  0.118 &  0.157 &  ----- \\
25\,M$_\odot$ 
&$^{ 19}$& F &  0.105 &  0.300 &  0.366 &  0.643 &  ----- \\
&$^{138}$&La &  0.106 &  0.192 &  0.901 &  1.604 &  2.244 \\
&$^{180}$&Ta &  1.382 &  2.360 &  4.238 &  6.238 &  7.102 \\
\end{tabular}
\end{ruledtabular}
\end{table}

\begin{table}
\centering
\caption{Production factor relative to solar normalized to $^{16}$O
production as a function of $\mu$ and $\tau$ neutrino temperature
(neutral current) and using $4\,$MeV for the electron
\mbox{(ani-)}neutrinos (for charged current only). ``Hax'' are the
results from \cite{woo90} using Haxton's $\nu$ cross sections and
``Kol'' for the new rates of this paper by Kolbe.
\label{table:nc}}
\begin{ruledtabular}
\begin{tabular}{r@{$\!\!\!\!\,$}lrrrrrrrrr}
&
& \multicolumn{4}{c}{$\hrulefill$ $15\,$M$_\odot$ $\hrulefill$}
& \multicolumn{4}{c}{$\hrulefill$ $25\,$M$_\odot$ $\hrulefill$}
\\
\multicolumn{2}{c}{product}
& \multicolumn{2}{c}{$\hrulefill$ $6\,$MeV $\hrulefill$}
& \multicolumn{2}{c}{$\hrulefill$ $8\,$MeV $\hrulefill$}
& \multicolumn{2}{c}{$\hrulefill$ $6\,$MeV $\hrulefill$}
& \multicolumn{2}{c}{$\hrulefill$ $8\,$MeV $\hrulefill$}
\\
&
& \multicolumn{1}{r}{Hax}
& \multicolumn{1}{r}{Kol} 
& \multicolumn{1}{r}{Hax}
& \multicolumn{1}{r}{Kol} 
& \multicolumn{1}{r}{Hax}
& \multicolumn{1}{r}{Kol} 
& \multicolumn{1}{r}{Hax}
& \multicolumn{1}{r}{Kol} 
\\
\noalign{\smallskip}\hline\noalign{\smallskip}
$^{ 11}$& B &  1.65 &  1.90 &  3.26 &  3.99 &  0.95 &  1.17 &  1.36 &  1.85 \\
$^{ 19}$& F &  0.83 &  0.64 &  1.28 &  0.82 &  0.56 &  0.37 &  1.03 &  0.55 \\
$^{ 15}$& N &  0.46 &  0.49 &  0.54 &  0.58 &  0.09 &  0.12 &  0.15 &  0.19 \\
$^{138}$&La & ----- &  0.97 & ----- &  1.10 & ----- &  0.90 & ----- &  1.03 \\
$^{180}$&Ta & ----- &  2.75 & ----- &  3.07 & ----- &  4.24 & ----- &  5.24 \\
\end{tabular}
\end{ruledtabular}
\end{table}

It was recently argued by Goriely et al.\cite{gor01}, following the
suggestion first made in \cite{woo90}, that charged current reactions
on $^{138}$Ba would dominate the production of $^{138}$La.  We confirm
this, finding the $\gamma$-process contribution is small and the
neutral current reaction on $^{139}$La is insignificant
(Figure~\ref{fig:la138}).  Sufficient $^{138}$La is made for about
solar co-production with $^{16}$O in massive stars
(Table~\ref{table:cc}).  The key is the enhancement of $^{138}$Ba by
the $s$-process prior to the SN (Figure~\ref{fig:la138prog}).
Conversely, charge-current production of $^{191}$Pt is insignificant 
because of the destruction of the parent $^{191}$Ir in the s-process.

The $\gamma$-process and neutral-current $\nu$-process each account
for about $25\,\%$ of $^{180}$Ta, with the remainder coming from the
charged current $\nu$ process.  It is somewhat overproduced.  However,
our calculations do not distinguish production in the $9^-$ excited
state from production in the $1^+$ ground state, while only the former
is long lived. Estimates \cite{rau02} indicated thermal freeze-out
should leave 30-50$\,\%$ in the isomeric state.  
A significant part of what is made by neutrino process from low-spin
parent nuclei, however, might preferentially cascade down to the $1^+$
state as the $\nu$ interaction is dominated by low multipoles
(A.~Hayes, priv. com.). This could account for the apparent
overproduction.

The $^{11}$B yield is somewhat higher than \cite{woo90}
(Table~\ref{table:cc}), but we note that this may also vary by up to a
factor 2 when modifying the $^{12}C$($\alpha$,$\gamma$) rate within
its current $30\,\%$ range of uncertainty \cite{kun02}.  $^{19}$F is
typically $50\,\%$ lower (Table~\ref{table:cc}) due to the reduced
cross section.  The neutral current cross section is uncertain as it
depends sensitively on the strong quenching of the GT strength as
predicted by the shell model and what fraction of the GT strength
resides above the particle threshold; experimental guidance is needed.
The neutrino contribution to $^{15}$N is increased by about $50\,\%$
using our new rates, but the total yield remains low.

The $^{138}$La and $^{180}$Ta yields are very sensitive to the
$\nu_e$ temperature, with $^{138}$La the better thermometer as
it is made almost exclusively in this channel (Tables~\ref{table:cc} and
\ref{table:nc}).  Such a thermometer is potentially quite important because
of the role of the as yet unmeasured neutrino mixing angle $\theta_{13}$.
This parameter, crucial to proposed terrestrial long-baseline neutrino 
searches for CP violation, will generate matter-enhanced $\nu_e \leftrightarrow
\nu_\tau$ oscillations at a density corresponding to the atmospheric
$\delta m^2$. Naively this corresponds roughly to the base of the carbon zone.
Such an oscillation would in turn ``heat'' the $\nu_e$s (assuming a
normal rather than inverted mass hierarchy), thus enhancing charged current rates.
As most of surviving $^{138}$La is produced at higher densities, in the outer
half of the neon zone, one might conclude that its production does not
probe this interesting physics.  However, it has been recently noted that
$\nu-\nu$ scattering may push the MSW resonance toward higher densities, thus into
the neon zone \cite{fuller}.  This is clearly an interesting question deserving
further study.    

{\smallskip\footnotesize AH, KL, GMP, and EK thank the INT for support
during the 2002 workshop ``nucleosynthesis''.  This research has been
supported by the NSF (AST 02-06111), the SciDAC Program of the DOE
(DE-FC02-01ER41176), the DOE ASCI Program (B347885), and by NASA
(NAGW-12036).  AH is supported in part by the DOE under grant B341495
to the FLASH Center at the University of Chicago and acknowledges
supported by a Fermi Fellowship of the Enrico Fermi Institute at The
University of Chicago. GMP is supported by the Spanish MCYT under
contracts AYA2002-04094-C03-02 and AYA2003-06128

}

\end{document}